\documentclass{article}

\usepackage{arxiv}

\usepackage[utf8]{inputenc} % allow utf-8 input
\usepackage[T1]{fontenc}    % use 8-bit T1 fonts
\usepackage{hyperref}       % hyperlinks
\usepackage{url}            % simple URL typesetting
\usepackage{booktabs}       % professional-quality tables
\usepackage{amsfonts}       % blackboard math symbols
\usepackage{nicefrac}       % compact symbols for 1/2, etc.
\usepackage{microtype}      % microtypography
\usepackage{lipsum}		% Can be removed after putting your text content
\usepackage{graphicx}
\usepackage{doi}

\usepackage{cite}
\usepackage[square, comma, sort&compress, numbers]{natbib}

\title{When Large Language Models Meet Optical Networks: Paving the Way for Automation}

\author{ 
    {Danshi Wang*, Yidi Wang, Xiaotian Jiang, Yao Zhang, Yue Pang, and Min Zhang}\\
    State Key Laboratory of Information Photonics and Optical Communications\\
Beijing University of Posts and Telecommunications, Beijing, China, 100876\\
	$^{\ast}$ Email:danshi\_wang@bupt.edu.cn \\
}

\renewcommand{\shorttitle}{When Large Language Models Meet Optical Networks: Paving the Way for Automation}

\begin{document}
\maketitle

\begin{abstract}
Since the advent of GPT, large language models (LLMs) have brought about revolutionary advancements in all walks of life. As a superior natural language processing (NLP) technology, LLMs have consistently achieved state-of-the-art performance on numerous areas. However, LLMs are considered to be general-purpose models for NLP tasks, which may encounter challenges when applied to complex tasks in specialized fields such as optical networks. In this study, we propose a framework of LLM-empowered optical networks, facilitating intelligent control of the physical layer and efficient interaction with the application layer through an LLM-driven agent (AI-Agent) deployed in the control layer. The AI-Agent can leverage external tools and extract domain knowledge from a comprehensive resource library specifically established for optical networks. This is achieved through user input and well-crafted prompts, enabling the generation of control instructions and result representations for autonomous operation and maintenance in optical networks. To improve LLM’s capability in professional fields and stimulate its potential on complex tasks, the details of performing prompt engineering, establishing domain knowledge library, and implementing complex tasks are illustrated in this study. Moreover, the proposed framework is verified on two typical tasks: network alarm analysis and network performance optimization. The good response accuracies and sematic similarities of 2,400 test situations exhibit the great potential of LLM in optical networks.
\end{abstract}

% keywords can be removed
% \keywords{Large language model \and Prompt engineering \and Domain resource library \and Optical network automation}

\section{Introduction}
Large Language Model (LLM) is a type of natural language processing (NLP) technology, which relies on deep learning with an extensive number of parameters and acquire abundant knowledge from large-scale datasets \cite{ref1}. The emergence of language models can be traced back to the 1950s and 1960s, when the foundations of NLP were initially established. Early endeavors in this field were propelled by rule-based systems and statistical methodologies. Before the advent of deep learning, Hidden Markov Models and N-gram models stood out as two popular approaches for NLP. In recent years, LLMs have achieved remarkable progress with the introduction of the Transformer architecture in 2017 \cite{ref2}.

Since then, numerous Transformer-based LLMs have been developed by several giant AI companies, such as Google’s T5 and Gemini, Meta’s LLaMa, Hugging Face’s BLOOM, and notably OpenAI’s GPT (Generative Pre-training Transformer). As a generative AI, GPT is currently considered as the most prominent and influential LLM, which can provide quick and reasonable answers to questions with human-like conversations and have achieved two major milestones: ChatGPT and GPT-4 \cite{ref3}. GPT has achieved state-of-the-art results on many NLP benchmarks and applications, such as intent understanding, content generating, text summarization, language translating, question answering, code debugging, and conversational robots in the fields of healthcare, education, law, finance, customer service, search engine, and scientific research. Typically, GPT is a generalist model that can effectively solve multiple general-purpose NLP tasks on a broad scale. However, it still faces challenges when tackling specific or complex tasks within particular professional domains \cite{ref4}. In such cases, it is imperative to devise suitable task instructions or formulate specific prompt strategies to elicit its specialized proficiencies.

As the cornerstone of informative infrastructures, optical networks are constructed with cross-connected optical fibers for massive data transmission between multiple nodes, with the characteristics of high capacity, long distance, large scale, and widespread deployment. Optical networks comprise lots of physical components, including optical transceivers, fiber optic cables, optical amplifiers, optical filters, optical switches, and other components. The categories of optical network can be classified as access network, metro network, and core network in terms of the service scope. In view of the fact that massive components, various topologies, and multiple scales, the operation and maintenance of optical networks are intricate and burdensome, including network monitoring, controlling, optimizing, upgrading, scheduling, troubleshooting, etc \cite{ref5}. Current network operation and maintenance heavily rely on human labors, involving a substantial workforce (domain specialist, network engineer, network administrator, network analyst, site worker, service personnel, and other staffs) engaged in specific and tedious tasks. Therefore, it is highly anticipated to develop an intelligent operation agent to assist network operators and engineers in diverse tasks\cite{refR4}, ultimately leading to the achievement of autonomous optical networks\cite{refR5}.

The past few years have witnessed a remarkable surge in research efforts from both academia and industry to seamlessly integrate and harness the potential of AI across various aspects of optical networks \cite{ref6}. However, previous studies have mainly relied on conventional machine learning or deep learning models with relatively small sizes. These small-sized models exhibit limited intelligence and can only perform a few specific functions, but falling short of achieving artificial general intelligence (AGI) and full automation capabilities. Meanwhile, extensive research on large models has shown that when scaling up the model size (over tens of billions of parameters) with augmented training data, the model capability can be significantly enhanced on a wide range of downstream tasks \cite{ref7}. The  advancement of LLM is paving the way for the realization of AGI and prompts us to expect the prospects of autonomous optical networks.

In this paper, encouraged by the leapfrog development and strong capabilities of LLMs, a framework of LLM-empowered optical networks is proposed. As the core component, an LLM-driven agent (AI-Agent) is deployed in the control layer, which can generate appropriate responses based on the user’s input, well-crafted prompts, and domain resource library. Through the generated control instructions and result representation, the AI-Agent facilitates autonomous control of the physical layer and efficient interaction with the application layer. To enhance LLM's specific capability in optical networks and maximize its potential on complex tasks, we have devised prompt engineering strategies and established a comprehensive domain resource library for the implementation of such tasks. Moreover, two typical tasks in optical networks (alarm analysis and autonomous optimization) are demonstrated as the case study of the proposed framework. The test results show that the application of LLM in optical networks has the great potential to simplify the operational processes and improve management automation. Furthermore, our work illustrates that LLMs can pave the way to a more intelligent, efficient and autonomous future in various fields, including optical networks.

The rest of the paper is organized as follows. In Section 2, we provide a detailed introduction to the framework of LLM-empowered optical networks, which enables intelligent control of the physical layer and efficient interaction with the application layer. In Section 3, the prompt engineering consisting of prompt elements and prompt techniques is introduced in detail. In Section 4, we describe the composition of domain resource library for optical networks and how to invoke tools and extract information from it. In Section 5, a five-step procedural framework for solving the complex tasks in optical networks is demonstrated. In Section 6, two case studies of alarm analysis and autonomous optimization are studied to illustrate the feasibility and effectiveness of LLM’s applications in optical networks. Finally, conclusions for this paper are provided in Section 7.

\begin{figure*}[t]
\centering\includegraphics[width=17cm]{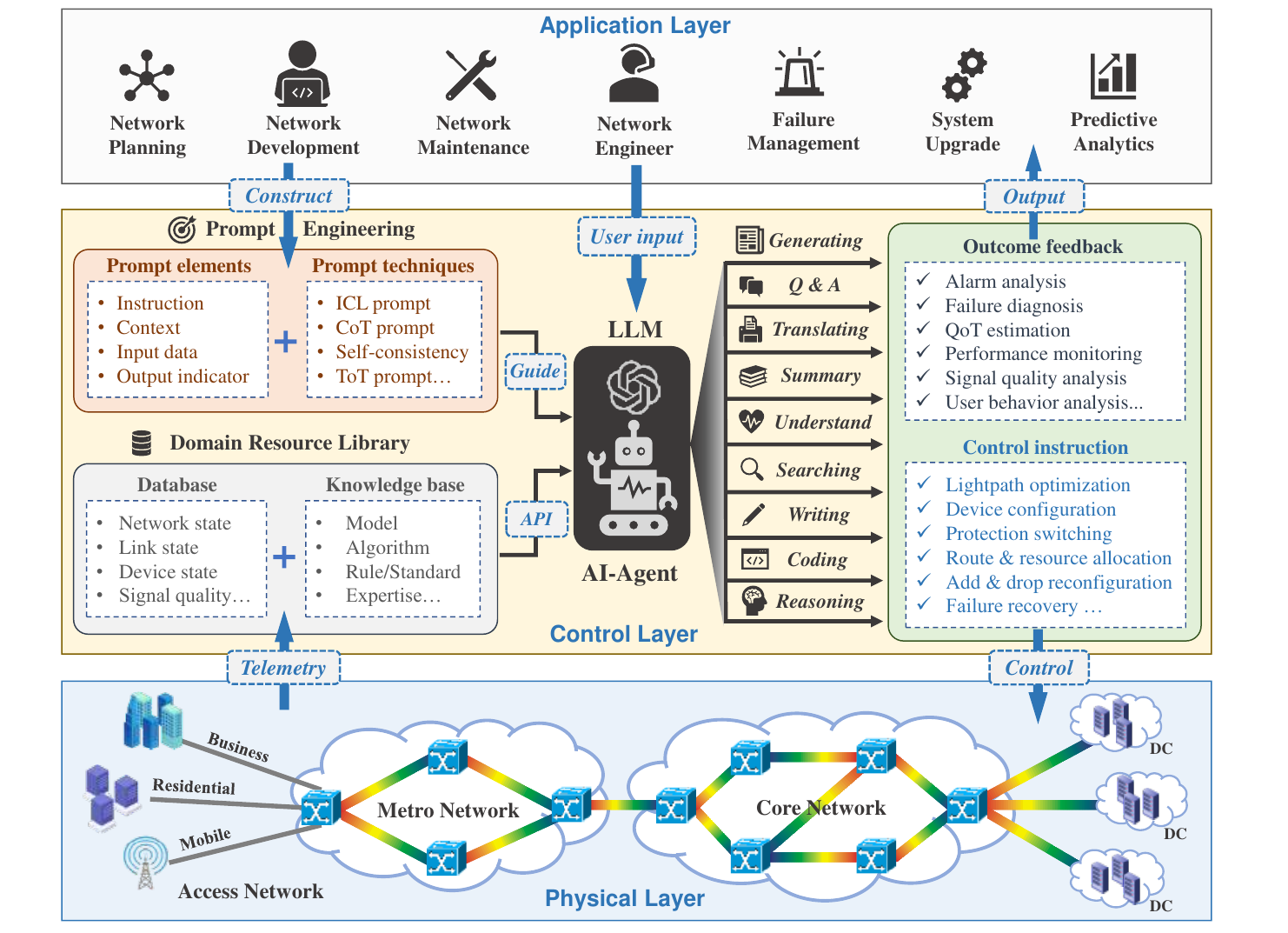}
\caption{The framework of LLM-empowered optical networks, which enables intelligent control of the physical layer and efficient interaction with the application layer through an LLM-driven AI-Agent deployed in the control layer.}
\label{img.1}
\end{figure*}

\section{Framework of LLM-empowered optical networks}
The proposed framework of LLM-empowered optical networks is illustrated in Fig. \ref{img.1}. The hierarchy of optical networks can be divided into access, metro, core, and data center (DC) networks. Moreover, optical networks are progressively advancing towards large-scale deployment, massive fiber-optic infrastructures, dynamic reconfiguration capabilities, ultra-wideband transmission, software-defined control mechanisms, resource virtualization, as well as open and disaggregated architectures. These emerging trends pose the challenges to the operation and maintenance of optical networks, including network planning, link optimization, hardware configuration, software development, failure management, system upgrade, etc.

To address these intractable issues, an LLM-driven intelligent assistant, termed “AI-Agent”, is designed and deployed in the control layer, as shown in Fig. \ref{img.1}. Its purpose is to assist network operators in facilitating intelligent control on the physical layer and efficient interaction with the application layer. In practical applications of such intelligent control, the AI-Agent is placed along with the software-defined network controller in the network layer. The LLM-generated commands are forwarded to physical layer devices through the Southbound Interface using standards such as the Network Configuration Protocol. In this paper, the GPT-series models are selected as the representative model for a case study. Different from conventional deep learning models that necessitate retraining for each task, GPT is a pre-trained LLM that has learned from extensive data that can directly perform various general-purpose tasks without the need for additional training. However, it always encounters challenges when applied to specific tasks in professional domains, and may fail to deal with complex issues in optical networks. In order to enhance professional capabilities of LLM for optical networks, it is imperative to augment LLM through the implementation of “\textit{Prompt Engineering}” and “\textit{Retrieval-augmented Generation (RAG)}”\cite{refR3}.

A “Prompt” is a sequence of tokens that serves as textual inputs to an LLM, instructing it to enhance its capability for a specific task. Prompt engineering is an emerging displine for developing and optimizing prompt to elicit the desired output from an LLM \cite{ref8}. It should be noted that for non-open source LLMs (e.g., ChatGPT and GPT-4), we have little to do in training or retraining phase. In such case, prompt engineering emerges as the most prominent step that we can do in developing LLMs for specialized applications. In optical networks, by crafting effective prompts, we can leverage the power and potential of LLMs for various purposes, such as answering operator’s queries for troubleshooting, understanding user’s intent, translating natural language into machine commands, generating control instructions for device configuration, summarizing useful information from massive network data, searching solutions from resource library, recording and reading network log, writing or debugging codes, reasoning failure cause, developing optimization strategy, etc. 

The RAG is a technique for enhancing the accuracy and reliability of LLMs with facts fetched from external sources, which references an authoritative domain resource library outside of its training data sources before generating a response \cite{ref9}. The domain resource library consists primarily of two parts: database and knowledge base. In optical networks, significant amounts of data are generated from a wide range of sources, including optical performance monitoring data, network condition data, device state data, alarm data, traffic data, and service data. These raw data can be collected in real-time through telemetry to control plane. All sorts of collected data comprehensively represent the network operating states and can be leveraged by the LLM-driven AI-Agent for information acquisition and state awareness. In addition, the field of optical networks have accumulated a wealth of knowledge base, encompassing diverse models, mathematical theories, physical mechanisms, operation rules, standard protocols, expertise, and other instructive information. These tools in knowledge base can be readily accessed by the AI-Agent through the application programming interface (API) to execute customized functions for the assigned tasks or provide expert knowledge for analysis and decision.

As described above, constructing effective prompts and leveraging appropriate domain resource can activate diverse potential capabilities of LLM. In optical networks, the LLM-driven AI-Agent is envisioned to play a pivotal role in autonomous network control and intuitive outcome feedback, facilitating efficient interaction across physical layer, control layer, and application layer, as depicted in Fig. \ref{img.1}. When implementing AI-Agent, the network operators can directly input their intents, tasks, or questions to AI-Agent in the manner of conversation. According to user’s input, AI-Agent will generate the corresponding response in the form of text, enabling the desired action through API integration and external tools. The control instruction that interacts with physical layer can be generated by AI-Agent to automatically execute tasks, such as lightpath optimization, device configuration, protection switching, failure recovery, route and resource allocation, impairment mitigation, etc. The outcome feedback that interact with application layer are output from AI-Agent to intuitively present the processed information and analysis results, such as alarm analysis, failure diagnosis, quality of transmission (QoT) estimation, performance monitoring, signal quality analysis, traffic prediction, and user behavior analysis. Therefore, in optical networks, the LLMs hold significant potential to accomplish various tasks for autonomous operation and active maintenance, while also facilitating a paradigm shift in human-network interaction.

\section{Prompt engineering for LLM}
Prompt engineering is an art aimed at crafting effective prompts that guide LLM to generate desired responses. As a fundamental technique in NLP, it plays a crucial role in maximizing the effectiveness of LLM by bridging the gap between user intent and model comprehension \cite{ref8}. To harness the full potential of LLM, a well-designed prompt is indispensable, which requires a good understanding and mastery of the basic elements and techniques of prompt engineering.

\subsection{Prompt elements}
To effectively convey the user’s intent to the LLM, a comprehensive and coherent prompt is generally required, consisting of four basic elements: instruction, context, input data, and output indicator. As the central element of prompt, instruction helps LLM construct logical reasoning chains by providing clear guidance on what the user wants, directly connecting the user intent to model comprehension. For further inference and analysis, necessary context would be helpful, which can be external information or additional background that guide the LLM to respond better. In more complex applications, instruction-related data is also part of the prompt, which drives the LLM to process the specific scenario user interests. Additionally, the output indicator is required when the LLM’s output is desired to be of a specific type or format. It should be noted that a well-designed prompt does not have to contain all the four elements. Depending on the complexity and requirements of the application scenario, the prompt may contain three, two, or even just the instruction element.

\begin{figure*}[t]
\centering\includegraphics[width=17cm]{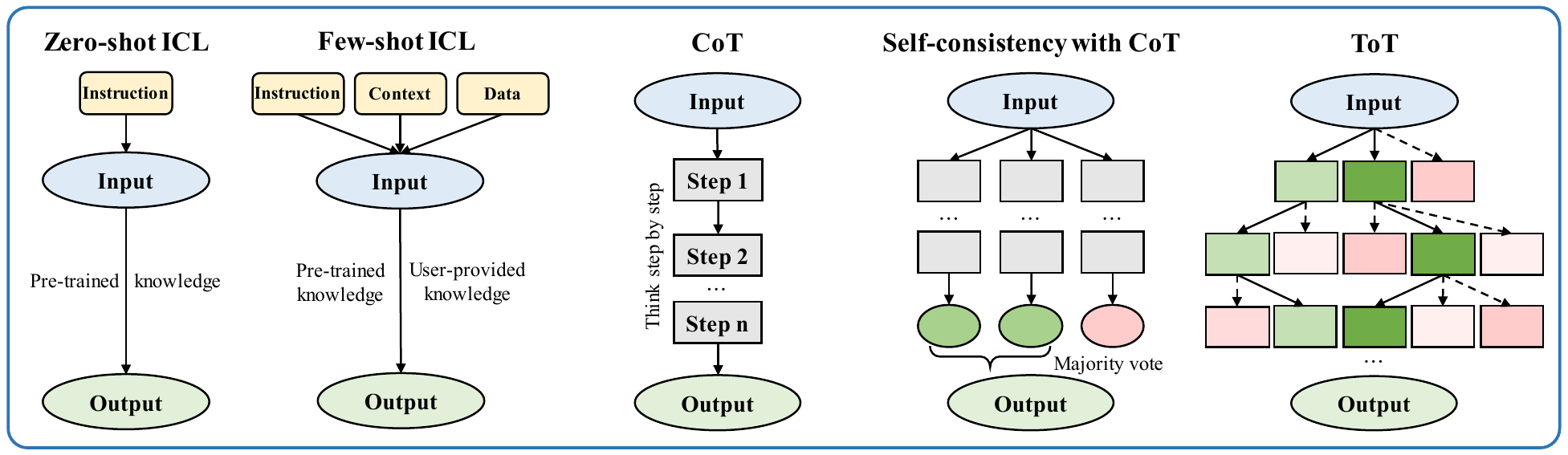}
\caption{Schematic of some typical prompting techniques, including zero-shot ICL, few-shot ICL, CoT, self-consistency with CoT, and ToT.}
\label{img.2}
\end{figure*}

\subsection{Prompt techniques}
In addition to prompt elements, another key point of prompt engineering is the prompt techniques, which plays a pivotal role in maximizing the information of prompt elements, and stimulating the LLM’s analyzing and reasoning potential to obtain high-quality output. The schematics of some typical prompting techniques are shown in Fig. \ref{img.2}. As the fundamental prompt technique, zero-shot in-context learning (ICL)\cite{refR1} is a simple and straightforward way to convey user’s intent to LLM by providing only instructions as input. However, it usually performs poorly in complex applications since the pre-trained knowledge can hardly satisfy the user-defined scenario. To address this problem, few-shot ICL emerges as a recommended solution, which yields promising results in intricate tasks by guiding the LLM learn from similar examples from domain resource library and input data. However, there still exists limitations in complex tasks using few-shot ICL, where multi-step reasoning and analyzing are required. By decomposing complex reasoning task into multiple simple subtasks, the chain of thought (CoT)\cite{refR2} prompting enables the LLM to think and reason step by step, thereby overcoming the limitation of insufficient reasoning ability. However, diverse reasoning paths may lead to different reasoning results. To ensure the output aligns with user’s intent, self-consistency is usually incorporated into CoT, which selects the most consistent answer as the final output according to the majority vote on different reasoning paths. Furthermore, the evaluation ability and search algorithm can also be introduced in each step of the chain, and the best step candidates are reserved for the next step of reasoning. Such tree of thoughts (ToT) prompting allows LLM to self-evaluate the progress that intermediate steps made towards completing the task through a deliberate reasoning process. Moreover, emerging prompting techniques such as automatic prompting, active prompting, and directional stimulus prompting have also been proposed recently for different types of tasks.

The successful implementation of LLM for desired tasks in optical networks heavily relies on the meticulous selection of prompt elements and seamless integration of prompt techniques, signifying the paramount significance of prompt engineering. Prompt engineering is an empirical science and the effect of prompt engineering methods can vary a lot among models, thus requiring heavy experimentation and heuristics. Consequently, it can be inferred that the future research on LLM will primarily strive to explore advanced techniques in prompt engineering.

\begin{figure*}[t]
\centering\includegraphics[width=17cm]{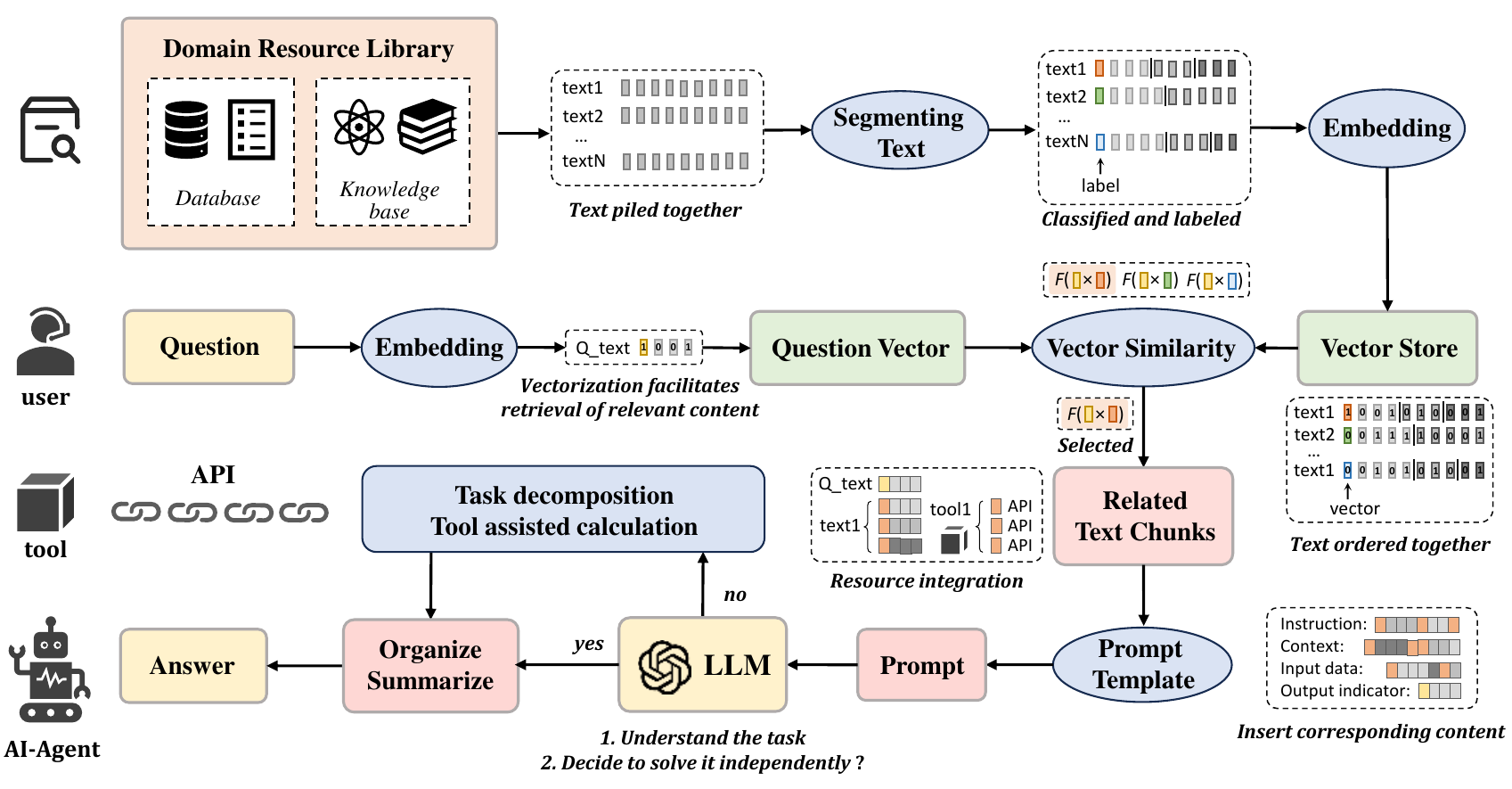}
\caption{Workflow of LLM to invoke tools and extract information from the domain resource library for executing specific tasks in optical networks.}
\label{img.3}
\end{figure*}

\section{Domain resource library}
In addition to prompt engineering, it is also significant to take full advantage of specialized domain resources for specific tasks in optical networks. To provide accurate and professional assistance in optical networks, it is essential to establish and maintain a comprehensive domain resource library. This library should be easily accessible and readily available for LLM, consisting of both database and knowledge base as depicted in Fig. \ref{img.3}. By extracting useful information from the database, it becomes convenient to search and provide tailored responses to specific problems, effectively activating its capability for targeted queries and tasks in optical networks; by invoking specialized model or expertise from the knowledge base, it can seamlessly integrate with LLM to create a more robust and versatile tool for targeted support and elaborate analysis.

\subsection{Data base}
In optical networks, significant amounts of data are generated from a wide range of sources, including historical data and real-time updated data that characterize the performance and states of equipment, link, and network. The equipment state data comprises measurements, such as gain, tilt, and noise figure from Erbium-doped fiber amplifier (EDFA), center wavelength, bandwidth, and attenuation value for each channel from wavelength selective switch (WSS), bit error rate (BER), launch and received power from optical module. The link performance data contains fiber loss profile obtained through optical time domain reflectometer (OTDR), optical power of each channel measured by optical channel monitor (OCM). The network condition data encompasses information regarding the topology, traffic, resource utilization, routing, alarms, and other relevant aspects. These data can be collected by telemetry through optical supervisor channel (OSC) and conveniently viewed through network management system.

\subsection{Knowledge base}
Over the past decades, the field of optical network has accumulated a wealth of professional knowledge, providing effective external tools for LLM. This domain knowledge base contains diverse models, algorithms, mechanisms, rules, protocols, and expertise \cite{ref10}. In the physical layer, the typical and widely-used tools include Gaussian noise (GN) model for QoT estimation, split-step Fourier method (SSFM) for fiber channel modeling, digital signal processing (DSP) algorithms in coherent detection for impairment compensation, EDFA and WSS configuration algorithms for system optimization. In the network layer, the effective and helpful tools include routing and wavelength assignment (RWA) algorithms, resource allocations strategies, traffic prediction algorithms, alarm analysis rules, fault management techniques, system upgrade operations, and network maintain guidance. All of these tools can be accessed through API between knowledge base and LLM.

\subsection{Invoking tools and extracting information from the domain resource library}
The domain resource library encompasses a variety of content types, including words, tables, numerical values, formulas, charts, documents, and codes, all of which can be represented in the form of text. Therefore, when establishing a local domain resource library, additional methods are required to systematically organize the above types of content, as shown in Fig. \ref{img.3}. First, the raw text data should be segmented into text chunks and ensure that each chunk is within the token limit. Then the vectors of segmented text chunks are extracted into a vector store by embedding method constructed by LangChain \cite{ref11} in parallel, aiming to facilitate finding the most relevant information as prompts of the potential to activate LLM.

The user’s query is transformed into a question vector for the purpose of indexing. By calculating the similarity between the question vector and the existing content in the vector store, we can retrieve and select one or more closely related text chunks based on their similarity. The extracted text chunks may include data, knowledge, algorithms, and models. These contents will be organized and presented in a sequential manner to create prompts for activating the specific capabilities of LLM, thereby generating corresponding responses. Inserting relevant content into the prompt template can help LLM better understand tasks and determine whether they can be directly solved with no need of task decomposition or external resources. According to the LLM’s response, the tasks will be categorized into two groups: simple tasks that can be solved directly and complex tasks that require further decomposed and tool invocation. Ultimately, all the answers with computational results are summarized into natural language feedback to the user, aligning with the operational requirements of the network and ensuring a professional user experience.

\begin{figure*}[t]
\centering\includegraphics[width=\textwidth]{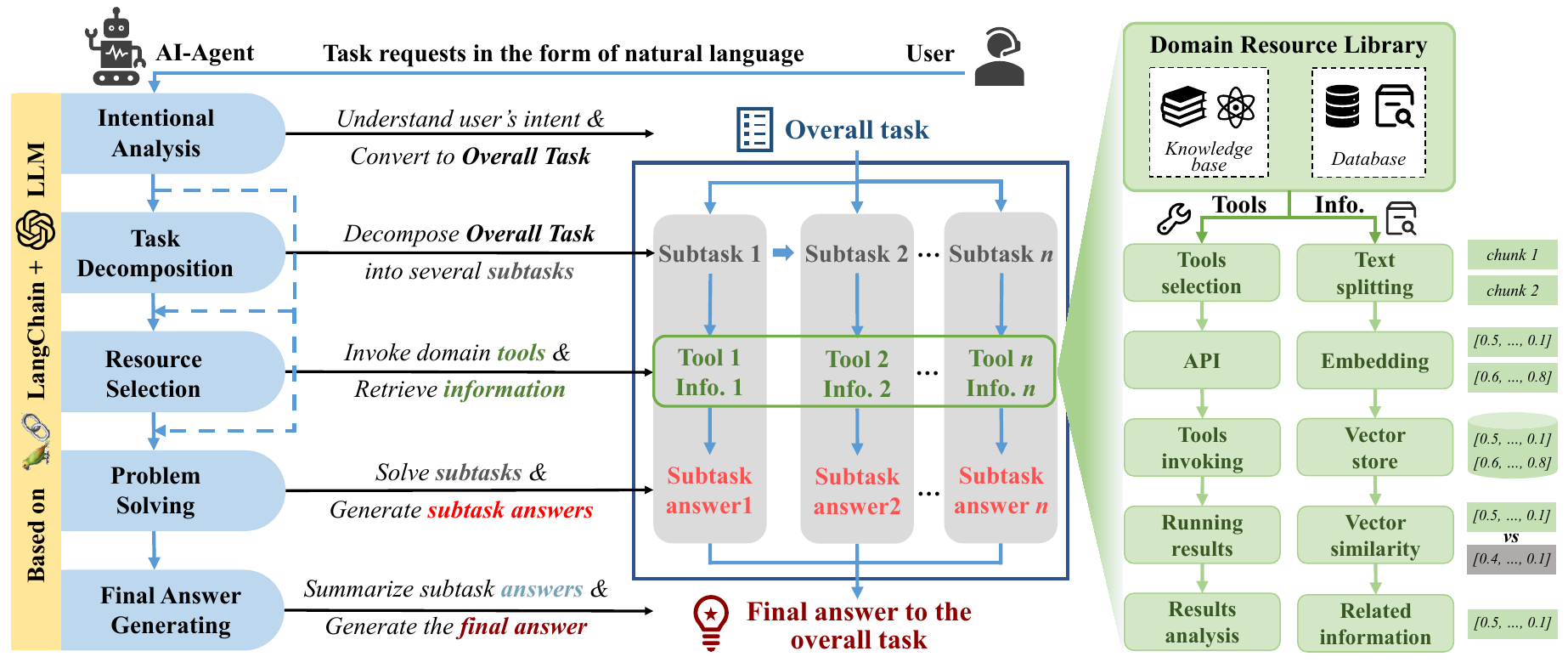}
\caption{Schematic of procedural framework for solving the complex tasks in optical networks based on LLM and LangChain, including five operating steps: intentional analysis, task decomposition, resource selection, problem solving, and final answer generating.}
\label{img.4}
\end{figure*}

\section{LLM for task implementation}
The operation and maintenance of optical networks involves a multitude of complex tasks that often encompass numerous factors and cannot be simply distilled into a singular problem, such as alarm analysis, fault diagnosis, and physical layer optimization \cite{ref12, ref13}. Instead, it usually necessitates the implementation of a systematic technique to solve the task in a step-by-step manner until achieving favorable outcomes. Here, we propose a five-step procedural framework for solving the complex tasks in optical networks based on LLM and LangChain \cite{ref11}, as depicted in Fig. \ref{img.4}:

\textbf{\textit{1) Intentional Analysis}}: As the first and most critical step in solving complex tasks, the AI-Agent should accurately understand the user’s intent and clarify the overall task. These functions are powered by a well-crafted LLM, which has a full understanding and learning of domain tasks and domain resources through mainly few-shot ICL. By learning typical task-solving cases, the agent can provide reliable intent analysis and overall task formulation for the specific task request from the user.

\textbf{\textit{2) Task Decomposition }}(optional): In most cases, the overall task cannot be answered perfectly in just one step, and it usually needs to be decomposed into several subtasks. Whether the overall task needs to be decomposed, how many subtasks it needs to be split into, and in which order are determined by the well-crafted LLM. With the help of prompt techniques, including CoT and ToT, the LLM could think logically like humans, and learns the chain-solving processes from a series of task-solving cases. For example, in order to analyze and handle the alarms in optical networks, the overall task is decomposed into three subtasks, including alarm compression, alarm correlation analysis, and alarm solving suggestions. By properly solving these three subtasks, the AI-Agent can generate better analysis and suggestions for alarms than those without task decomposition.  

\textbf{\textit{3) Resource Selection }}(optional): Generally, it is tricky for LLMs to solve the diverse task requirements in professional domains, as they often necessitate capabilities beyond NLP. The advanced power of LLMs in professional domains will emerge when they are combined with the domain-specific resources. On the one hand, the LLM can access and invoke domain tools through API based on its tool selection decisions, and subsequently analyzes the obtained results. On the other hand, the professional information in the domain database can be retrieved quickly and accurately as required, thanks to the RAG technique integrated in LangChain. Specifically, different types of knowledge and data are first loaded and split into smaller chunks. Each chunk is further embedded to capture its semantic meaning so that efficiently identifying similar chunks. The embedded chunks are then stored in the vector store, and a similarity analysis is executed to select the most related chunks for further analysis.

\textbf{\textit{4) Problem Solving}}: In this step, the LLM will generate a comprehensive answer for each subtask by leveraging the results obtained from the invoked tools, the relevant information retrieved from domain database, and the self-analysis of LLM pertaining to both current and previous subtasks. The generated answers can effectively solve the current subtasks and provide valuable information for tackling the next subtask. Therefore, similar to the approach employed by experts and experienced engineers in handling complex domain tasks, LLM is able to generate better results by systematically solving subtasks step by step.

\textbf{\textit{5) Generating Final Answer}}: Finally, the well-crafted LLM will generate a final answer to the overall task, generally including two patterns: (a) the cascaded pattern, where the outcome of the previous subtask will have an impact on the subsequent one and the final answer is derived from the result of the last subtask; (b) the parallel pattern, where each subtask is independently and simultaneously solved, and the final answer is obtained from a comprehensive summary of all subtasks’ results. This approach ensures the final answer is presented with clarity and completeness, thereby enhancing the overall effectiveness of the problem-solving process.

In the process of analysing and solving numerous problems in optical networks, these five steps can ensure the accuracy, professionalism, and comprehensibility of answers, effectively assisting network operators in tackling complex problems. Note that as supporting technologies, both LLM itself and LangChain are constantly applied throughout the entire process to facilitate effective operation at each step.

\section{Applications of LLM in optical networks}
Next, we will apply the LLM to optical networks based on above-introduced techniques. Herein, we study two typical tasks: network alarm analysis and autonomous network optimization. In this section, the up-to-date GPT-4 is selected as the representative LLM for the case study.

\subsection{Network alarm analysis}
Network alarms are the signs that indicate the link status or device incident, such as fault, anomaly, or degradation. Alarm analysis involves the process of detecting, diagnosing, locating, and recovering network issues by analyzing massive alarm data, which is significant for network operators but still faces challenges in terms of low efficiency, strong reliance on expertise, strenuous manual labor.

\begin{figure*}[t]
\centering\includegraphics[width=17cm]{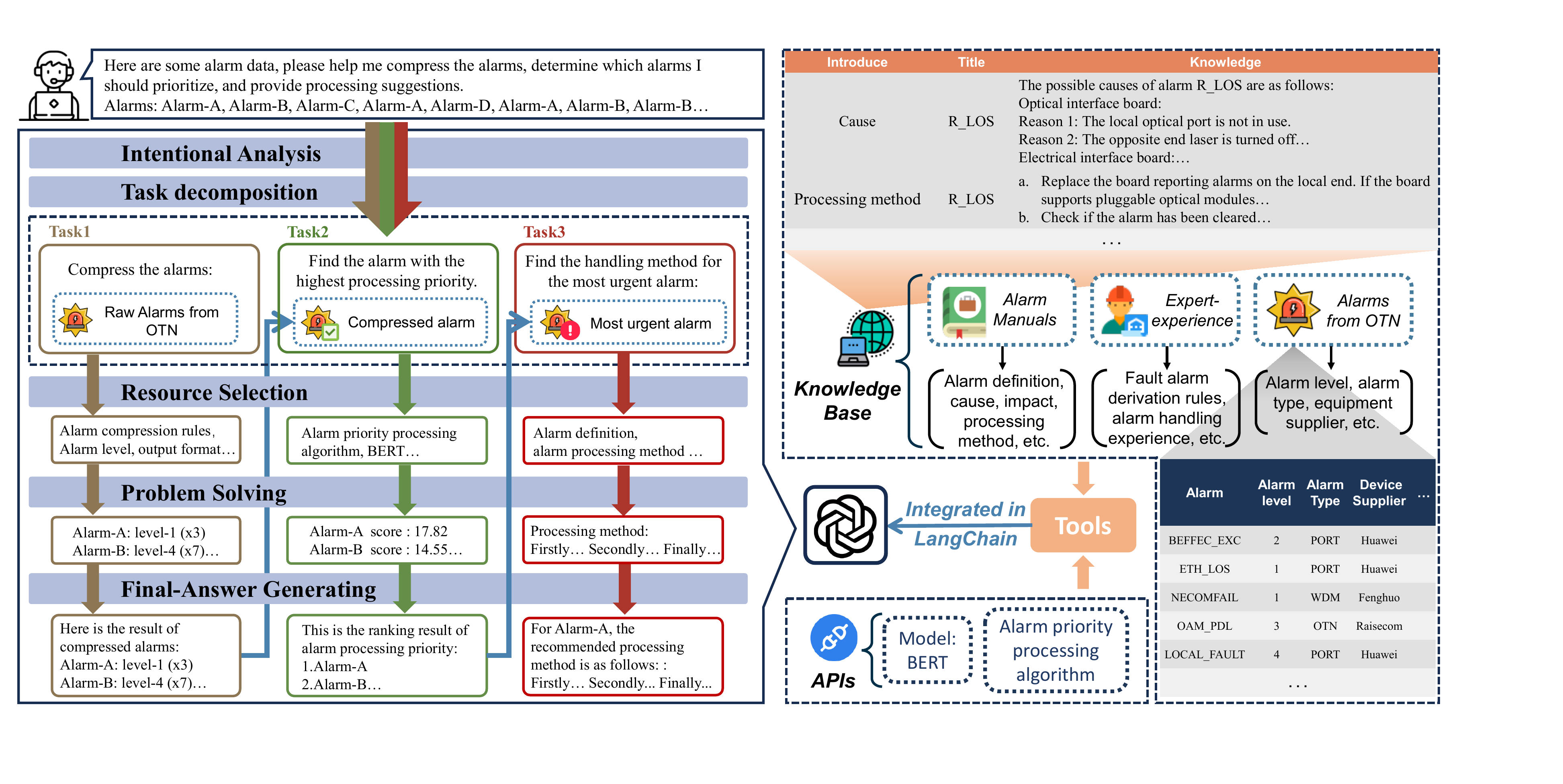}
\caption{Schematic of LLM-enabled network alarm analysis based on five-step procedural framework, encompassing three tasks: alarm compression, process priority sorting, and alarm solving suggestions.}
\label{img.5}
\end{figure*}

We combine network alarm analysis with LLM to automate a series of processes, including alarm compression, priority sorting, and providing alarm handling suggestions \cite{ref14}, as shown in Fig. \ref{img.5}. The knowledge extracted from the alarm manual, expert experience, and information from real OTN alarms are integrated into the knowledge base. BERT and alarm priority processing algorithms are integrated into the database. The knowledge base and database are integrated as tools in LangChain and can be called by LLM. 

Compared to directly querying GPT, our method allows GPT to better understand the principles of alarm analysis and proceed to analyze and solve problems step by step. Then the overall question is directly provided to GPT-4. After analyzing the operator's intention, the overall question will be divided into three tasks, each of which will be resource selected, task solved, and finally an answer will be generated.

\begin{figure*}[t]
\centering\includegraphics[width=12cm]{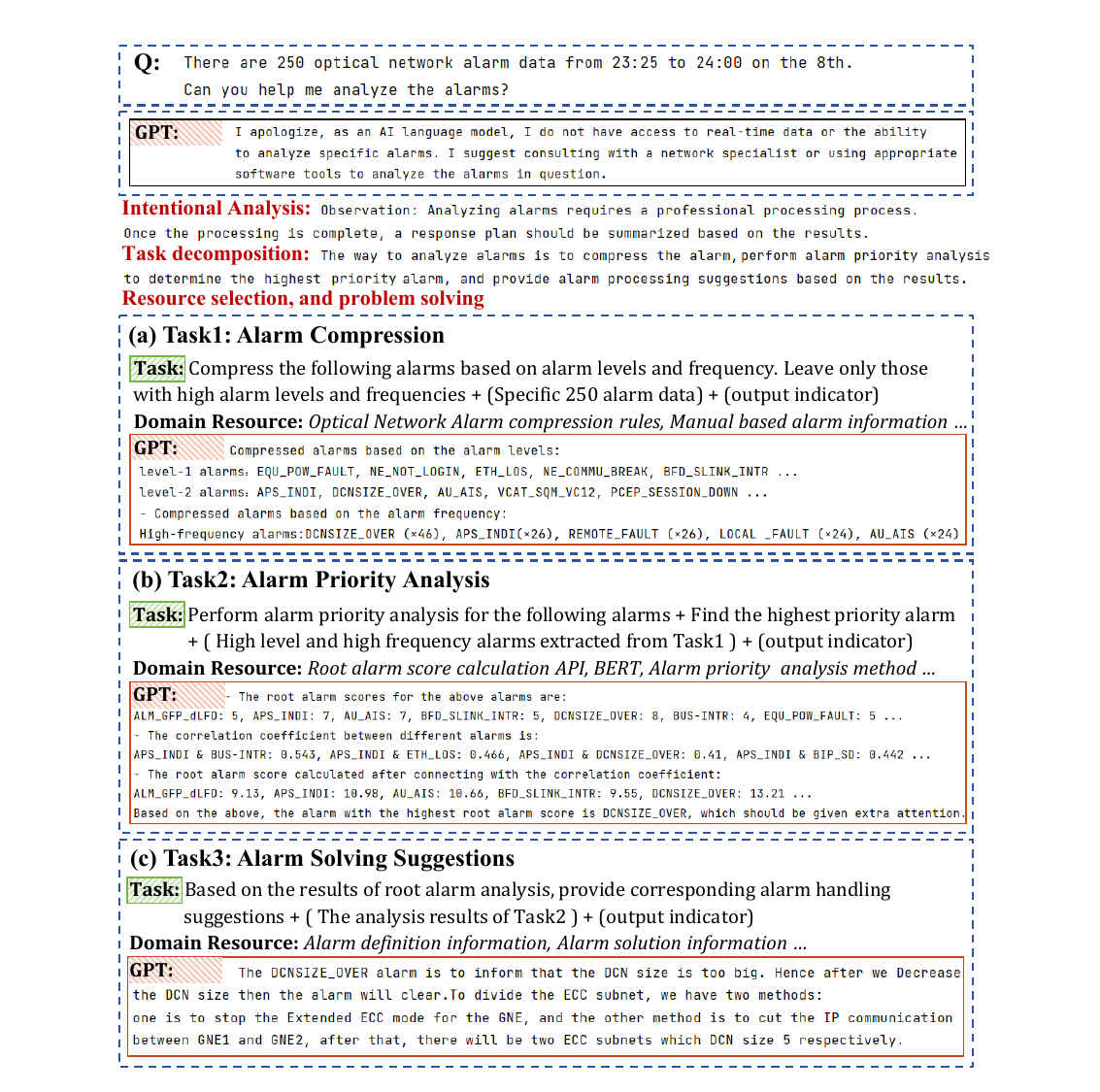}
\caption{Results of network alarm analysis using LLM, including alarm compression, alarm correlation analysis, and alarm solving suggestions.}
\label{img.6}
\end{figure*}

\textbf{\textit{Task1: Alarm compression.}} The urgency of alarms in OTN generally depends on two aspects: the severity level and occurrence frequency. The key to alarm compression lies in these two features, and the compressed alarm can help operators visually observe the overall situation of the alarm event. The OTN alarm information in the knowledge base can help LLM achieve more accurate alarm compression.The actual effect of alarm compression is shown in Fig. \ref{img.6} (a).

\textbf{\textit{Task2: Process priority sorting.}} This task aims to calculate and update the alarm importance score based on the compression results in Task 1. After LLM obtains the compressed alarm event information, it obtains the importance score of each alarm by calling BERT and alarm priority processing algorithms. Among them, BERT can obtain the correlation coefficient between different alarms, and the alarm priority processing algorithm combines the alarm level, occurrence frequency, and the correlation coefficient between alarms to calculate the importance score. Then LLM can sort the priority of alarm processing based on the results.The actual effect of alarm priority analysis is shown in Fig. \ref{img.6} (b).

\textbf{\textit{Task3: Alarm solving suggestions.}} The alarm with the highest score in Task 2 will be identified as the most urgent alarm of this event. LLM will provide the cause and handling suggestions for this alarm based on the information from the alarm manual and guidance from the alarm processing rule, as the conclusive response generated by AI-Agent.The actual effect of providing alarm suggestions is shown in Fig. \ref{img.6} (c).

With the help of carefully prepared CoT prompts, LLM is activated to understand and break down the problem, select appropriate knowledge and APIs, and solve the problem step by step. This comprehensive alarm processing scheme showcased the capabilities of LLM, providing a successful solution to this intricate task in optical networks.

\subsection{Autonomous network optimization}
Next, we study another LLM-based scheme for autonomous network optimization with a sequence of subtasks, including QoT estimation, network analysis, resource allocation, and step-by-step performance optimization \cite{ref15}. 

\begin{figure*}[t]
\centering\includegraphics[width=17cm]{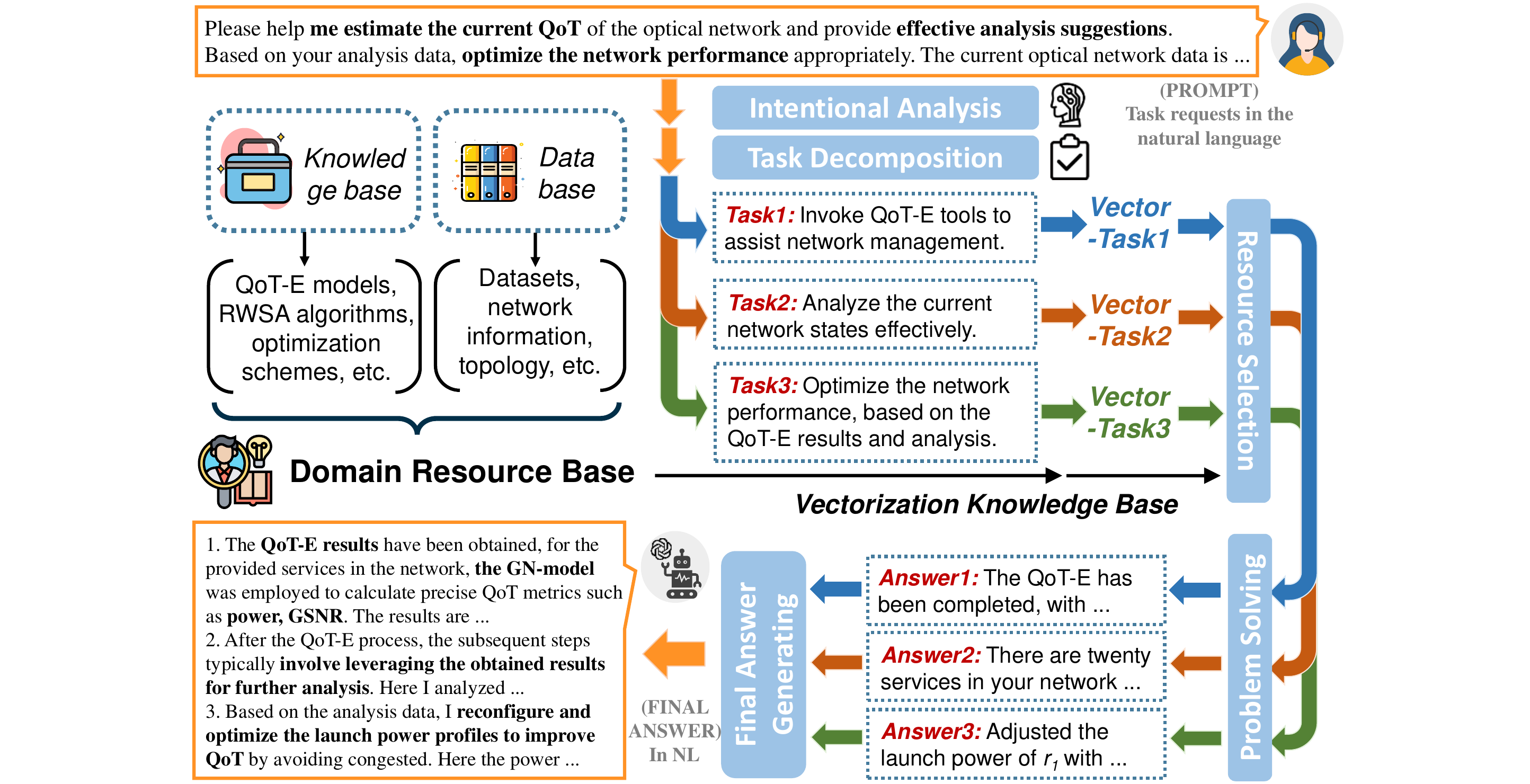}
\caption{Schematic of LLM-enabled autonomous network optimization based on five-step procedural framework, encompassing three tasks: QoT estimation, network analysis and suggestion, and performance optimization.}
\label{img.7}
\end{figure*}

We pass on the requests of automated QoT estimation and network performance optimization and its background to GPT-4 through prompts. The network topology information, link physical parameters, and system configuration rules required for the implementation of the tasks are used as knowledge, while network configuration interfaces, computing tools, optimization algorithms, and others are organized into the resource library. The scenario we validated is in the CONUS network with 77 nodes and 99 bidirectional connections, transmitted through the C+L-band. We provided initial information for 15 services distributed throughout the network that require QoT estimations, including the source and destination nodes of the services, launch power and GSNR profiles. When a question is submitted to the AI-Agent, the initial problem is decomposed into three subtasks that are matched with related knowledge from the domain resource library in the form of vectors to be completed in sequence, as shown in Fig. \ref{img.7}.

\begin{figure*}[t]
\centering\includegraphics[width=12cm]{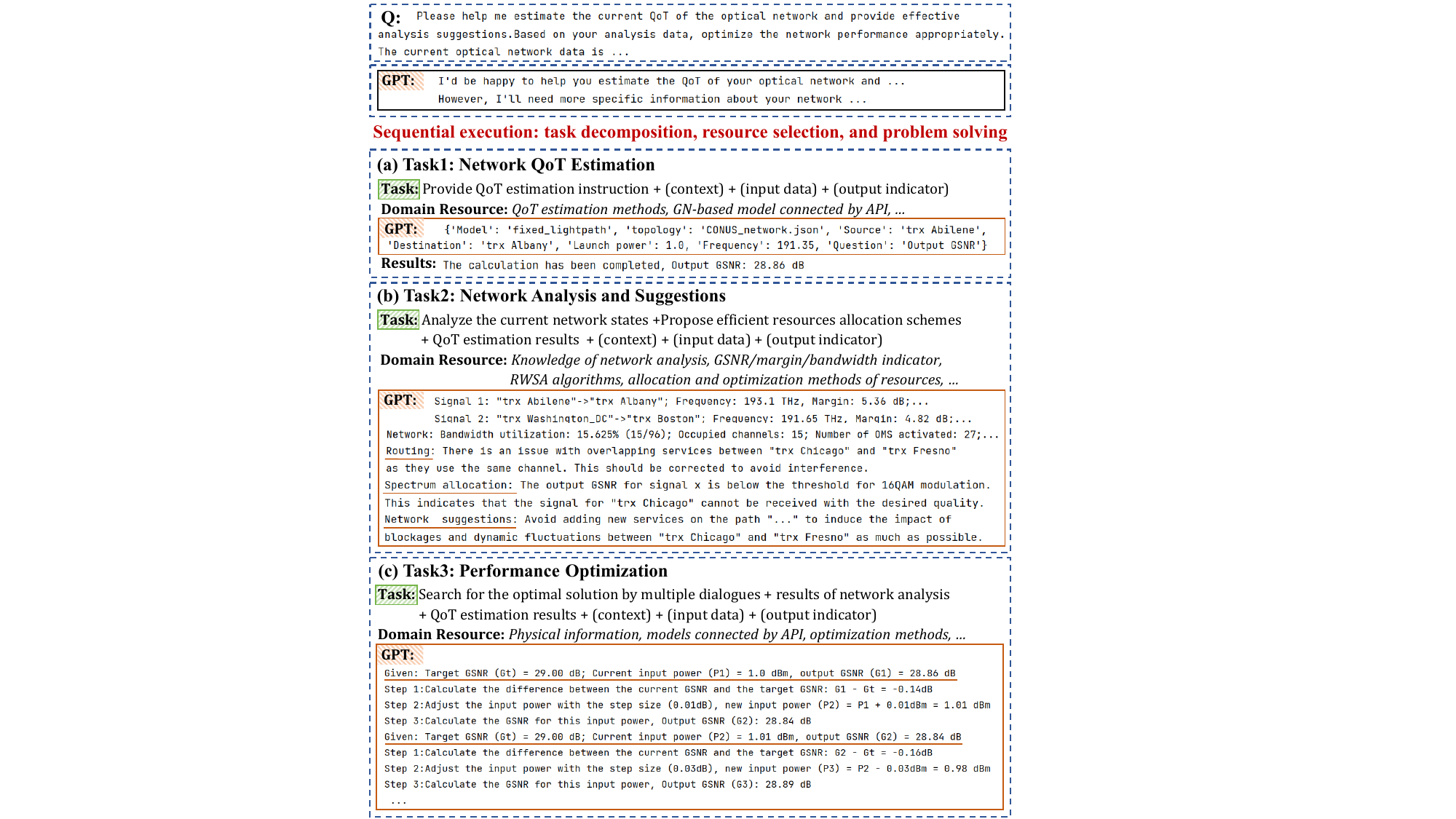}
\caption{Results of autonomous network optimization using LLM, including QoT estimation, network analysis and suggestions, performance optimization.}
\label{img.8}
\end{figure*}

\textbf{\textit{Task1: QoT estimation.}} It is the process of predicting and evaluating the QoT in terms of GSNR. LLM assists in transferring instructions based on the basic network information such as network topology files, elements configuration files, and parameters monitored; as well as user-provided input parameters, such as signal power, modulation format, and frequency. In this step, natural language in prompt is transformed into computer instructions, and the GN model \cite{ref16} in knowledge base are invoked for precise QoT estimation. This prediction serves as a foundation for subsequent network analysis. Fig. \ref{img.8} (a) shows the domain resources provided during the execution of this task. GPT translates our intention into the instructions and we show the final calculation result.

\textbf{\textit{Task2: Network analysis and suggestions.}} In this step, the results of QoT estimation and data for network running are utilized as prompt to enable LLM to conduct comprehensive network analysis. The proposed scheme evaluates a range of network metrics, as shown in Fig. \ref{img.8} (b), including power, GSNR and margin of each channel after each link; blocking probability, spectrum resource utilization and other relevant factors calculated through formulas based on knowledge to comprehensively assess the overall network performance and identify potential parts for improvement. 

\textbf{\textit{Task3: Performance optimization.}} Based on the resource allocation recommendations, simultaneously in the third step, LLM guides the iterative optimization process step-by-step. It assists in adjusting launch power profiles and EDFA configuration \cite{ref16} for the GSNR optimization. This iterative process fine-tunes network parameters, ensuring that optimal transmission conditions are achieved for each connection. Fig. \ref{img.8} (c) shows the output of the iterative process during optimization.

Then the final answer for the initial question consisting of three subtasks can be obtained from AI-Agent. The proposed scheme showcases the potential of leveraging LLM in optical networks. By sequentially addressing QoT estimation, network analysis, resource allocation, and GSNR optimization, the LLM can autonomously enhance network performance.

\subsection{Results and analysis}
\begin{figure*}[t]
\centering\includegraphics[width=17cm]{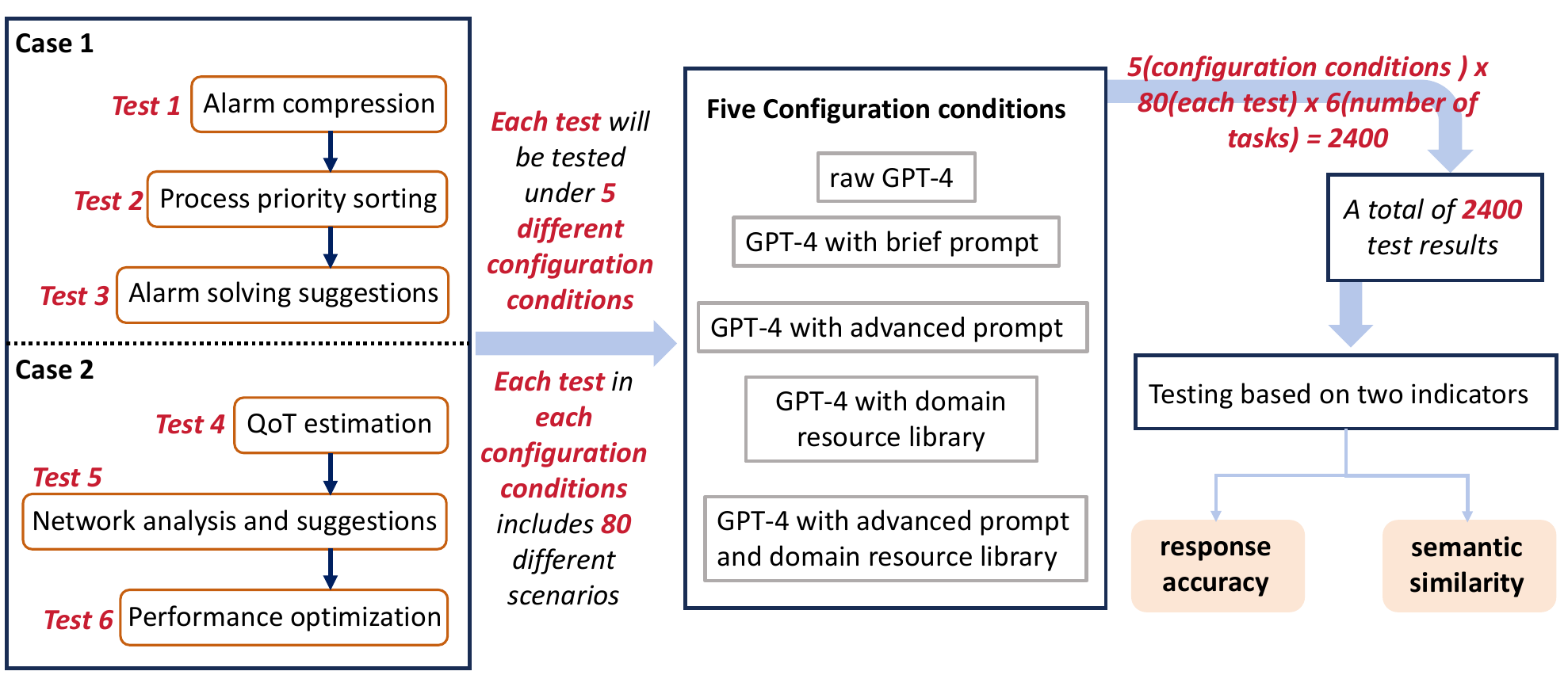}
\caption{Implementation flowchart for result testing and analysis.}
\label{img.a}
\end{figure*}

In the professional domain, LLMs sometimes suffer from confabulations (or hallucinations) which can result in them making plausible but incorrect statements in some tasks. To study the feasibility and efficacy of LLM in optical networks, we conducted extensive validations on the two case studies encompassing six tasks, focusing on response accuracy and semantic similarity. Total of 2,400 situations were tested (400 situations for each task) under five configuration conditions. More detailed information about the tests used, the evaluation methods, and the processes is presented in Fig. \ref{img.a}.

As shown in the Fig. \ref{img.a}, the test utilized two cases, each containing three types of test content. For the network alarm analysis case, these included alarm compression, processing priority analysis, and alarm resolution suggestions. For the autonomous network optimization case, these included QoT estimation, network analysis and suggestions, and performance optimization. In total, there are six types of test content, which need to be tested in five different configuration conditions (including raw GPT-4, GPT-4 with brief prompt, GPT-4 with advanced prompt, GPT-4 with domain resource library, and GPT-4 with both advanced prompt and domain resource library, where ICL and CoT are applied in the advanced prompt). Each test includes 80 scenarios in each configuration condition, resulting in a total of 2400 scenarios across all tests. These 2400 test results are evaluated using two metrics: response accuracy and semantic similarity.

In the network alarm analysis test, 2,500 out of 2,9541 alarms generated in a real OTN within 2 days were adopted, and each 25 adjacent alarms were packaged and randomly selected for testing within a time window of approximately 3 minutes. In autonomous network optimization, we randomly generate a set of traffic information each time under the same CONUS network topology and link parameters to verify the QoT estimation. Random network state information containing anomalies was provided in the network analysis task to test whether LLM can accurately analyze insufficient margin, service congestion, conflicts, etc. In the optimization task, we provide random network states to be optimized and verify the optimization task of the network by invoking tools. 

\begin{figure*}[t]
\centering\includegraphics[width=17cm]{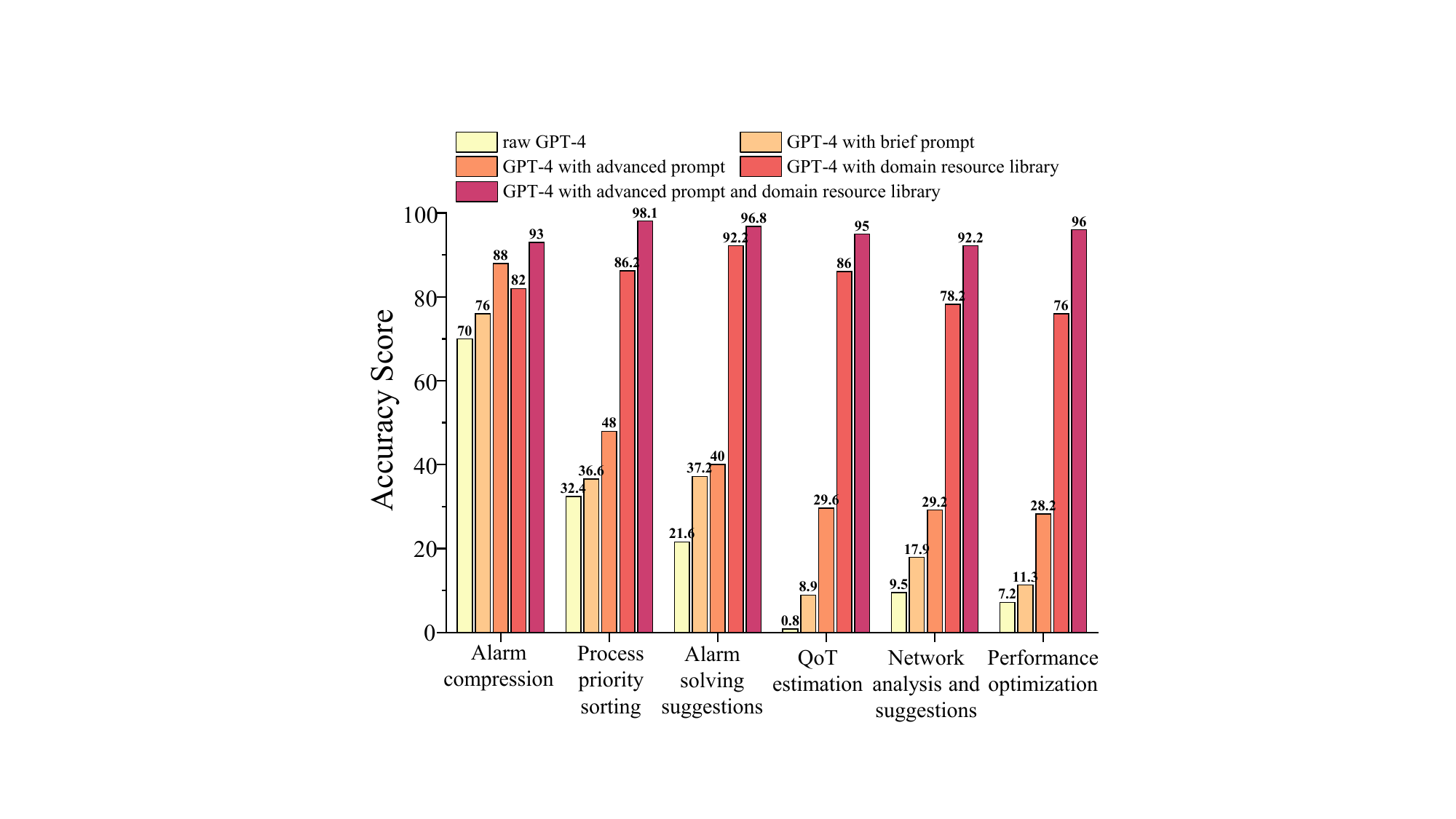}
\caption{Response accuracy of six tasks for network alarm analysis and autonomous network optimization under five configuration conditions.}
\label{img.9}
\end{figure*}

\begin{figure*}[h!]
\centering\includegraphics[width=17cm]{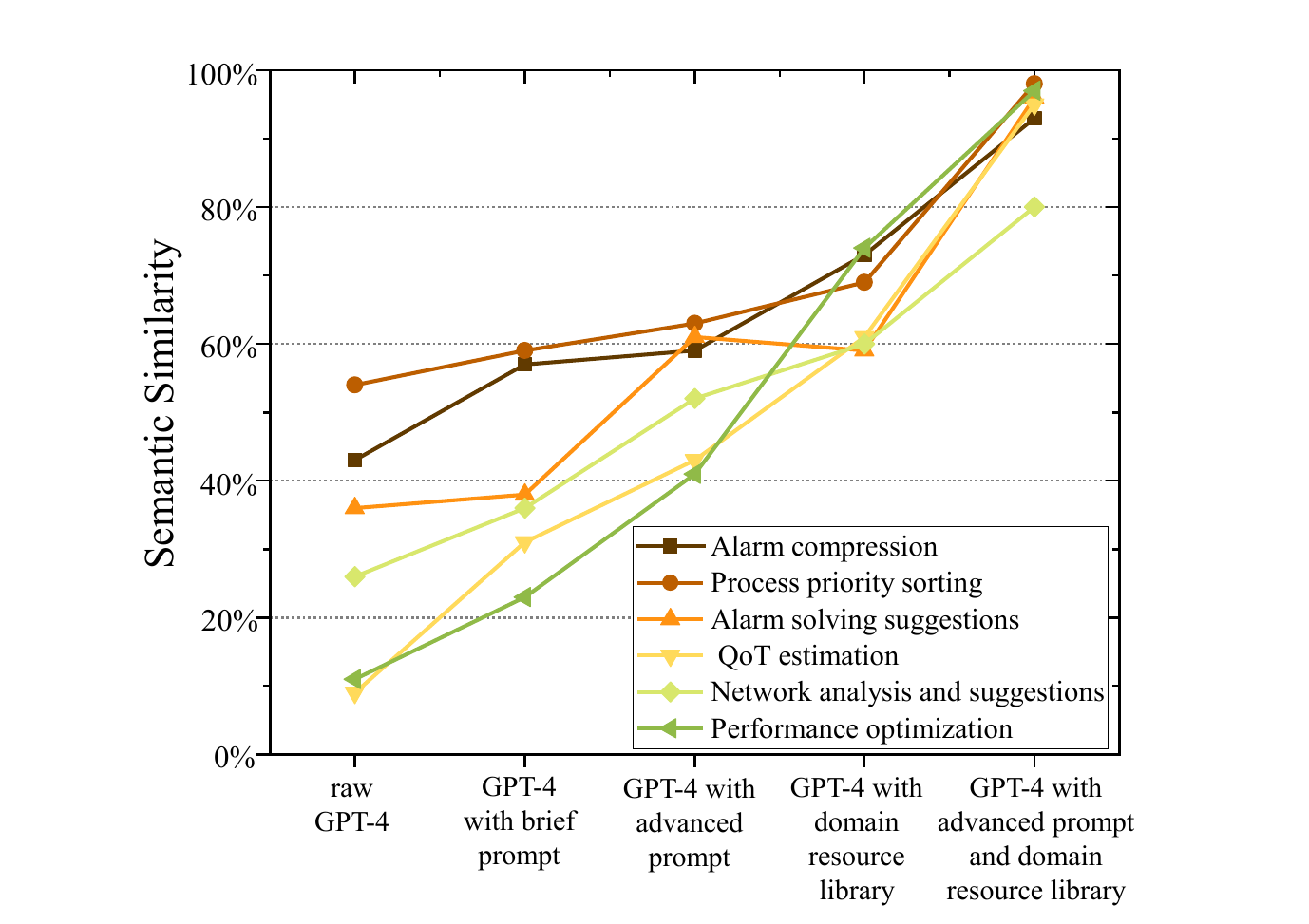}
\caption{Semantic similarity of six tasks in network alarm analysis and    autonomous network optimization under five configuration conditions.}
\label{img.10}
\end{figure*}

First, we evaluated the response accuracy of AI-Agent on each task under five different configuration conditions to investigate the impact of well-crafted prompts and domain resource library on performance variations in different tasks. Since some open tasks (such as alarm solving suggestions, network analysis and suggestions) are highly subjective, we formulate a specific scoring rule to evaluate the answers generated by LLM. Each answer is scored based on the proportion of key elements it contains, including validity of information, completeness of each step, accuracy of intermediate results, consistence of format, and expert judgement \cite{ref17}. Under this scoring rule, the accuracies of LLM on these six tasks under different configuration conditions are calculated, as shown in Fig. \ref{img.9}. In all tasks, the raw GPT-4 exhibits the poorest performance, especially in the three tasks of autonomous network optimization that rely more heavily on domain resources. Although only brief prompts are added, the accuracy is improved to a certain extent, and more accurate results can be achieved when advanced prompt is applied, illustrating that well-crafted prompts are significant to improve the LLM’s performance. Except for alarm compression, significant accuracy improvements are achieved in other tasks with the help of domain resource library, which illustrates that domain resources are indispensable for domain-specific problems. Particularly, the exception in alarm compression can be primarily attributed to the fact that alarm compression is essentially a traditional NLP problem based on the severity level and occurrence frequency, which requires much fewer domain resources than other tasks. When both advanced prompt and domain resource library are designed and established, the AI-Agent achieves the optimal accuracy in all tasks, thereby demonstrating the effectiveness of combining prompt engineering and domain resource library for complex tasks in professional domain.

Moreover, we analyze the semantic similarity between the answers generated by the LLM and the expected outputs. The purpose of this validation is to verify the effectiveness of the generated results as effective instructions for interaction among application layer, network layer, and physical layer. The SentenceTransformer \cite{ref18} is applied as the evaluation model, because it is capable of generating meaningful embeddings that reflect semantic content, enabling semantic similarity assessment to ensure reliability and reproducibility. Similar to the accuracy validation experiment, the three subtasks of network alarm analysis achieve higher semantic similarity than the tasks in autonomous network optimization when only prompt engineering is applied, as shown in Fig. \ref{img.10}. Compared with only applying the prompt engineering, when only the domain resource library is applied, the performance optimization task achieves a great improvement, while the alarm solving suggestions task even experienced a decrease of 2\%. After integrating the advanced prompt with the domain resource library, the similarity of all six tasks is significantly improved than utilizing only one of them, proving that the combination of these two technologies plays a pivotal role in standardizing the responses generated by the LLM.

In the optical network framework empowered by LLMs, despite its significant advantages in intelligent physical layer control and efficient interaction with the application layer, potential ethical challenges and security risks must be addressed. Therefore, we propose several solutions to enhance the system's security and reliability. First, automatic monitoring and verification modules, as well as human evaluation, can be implemented to assess the quality of LLM-generated responses. Low-quality or potentially harmful responses should be discarded to ensure the safe operation of the network. Additionally, a reliable logging system to record operations and outputs at each stage is necessary, as it can effectively help operators quickly identify the root cause of issues, improving the accuracy of commands and network security. Lastly, the access and operation permissions of the LLM on the network can be restricted. Considering network security, for operations that may significantly impact the physical layer or system, the LLM's outputs should serve as recommendations for operators rather than direct commands. Manual verification and review must be conducted before any instructions are issued by the LLM to further ensure their correctness. Network administrators should check the accuracy and applicability of the instructions based on their operational experience to prevent unexpected interruptions or failures. Through the above measures, we can effectively reduce the potential security risks and ethical issues while fully leveraging the advantages brought by LLM technology, ensuring the safe and reliable operation of optical networks.

\section{Conclusions}
In this study, we proposed a framework of LLM-empowered optical networks that enabled intelligent control of the physical layer and efficient interaction with the application layer through a AI-Agent deployed in the control layer. According to the user’s input and well-crafted prompts, the AI-Agent can accordingly invoke external tools and extract domain knowledge from the established domain resource library to generate control instructions and result representations, thereby realizing intelligent network operation and maintenance. We extensively deliberated on strategies for proficient prompt engineering, establishment of a comprehensive domain resource library, and successful implementation of intricate tasks, all of which are indispensable for enhancing LLM's proficiency in professional domains and unlocking its potential in tackling complex assignments. Moreover, the applications of LLM in network alarm analysis and autonomous network optimization were also studied. The results showed that LLM exhibited expertise in optical networking. Predictably, LLMs will play an increasingly important role in optical networks in the future, fundamentally transforming the traditional operation and maintenance paradigm.

\end{document}